\PassOptionsToPackage{pdfpagelabels=false}{hyperref}
\documentclass[useAMS,usenatbib, usedcolumn]{mnras}
\pdfoutput=1 
\usepackage{graphicx}
\usepackage{color}
\usepackage{amsmath,amssymb,amstext}
\usepackage[T1]{fontenc}
\usepackage{mathptmx,txfonts}
\usepackage[figure,figure*]{hypcap}
\usepackage[most]{tcolorbox}
\usepackage{ae,aecompl}
\input{hyperlink-year-only-natbib-patch}

\hypersetup{
    pdfauthor={Victor F. Calderon},
    pdftitle={Small and Large-Scale Galactic Conformity in SDSS DR7},
    pdfkeywords={cosmology: observations --- cosmology: large-scale structure
            of Universe --- galaxies: groups --- galaxies:clusters
            --- methods: statistical},
    colorlinks=true,
    citecolor=blue,
    linkcolor=magenta,
    urlcolor=cyan}

\usepackage[utf8]{inputenc}
\usepackage[english]{babel}
\usepackage{array}
\usepackage{afterpage}
\usepackage{color}
\usepackage{graphicx}
\usepackage{amsmath, amssymb}
\usepackage{soul,xcolor}
\usepackage{epstopdf}
\usepackage{units}
\usepackage{relsize}
\usepackage{enumitem}
\usepackage{url}
\usepackage{appendix}
\usepackage{footnote}
\usepackage{xspace}
\usepackage{float}

\usepackage{booktabs}
\usepackage{tabularx}

\usepackage{color}
\usepackage{graphicx}
\usepackage{epstopdf}
\usepackage{enumitem}
\usepackage{xspace}
\usepackage{float}
\usepackage{booktabs}
\usepackage{tabularx}

\usepackage{relsize}
\usepackage{array}
\usepackage{afterpage}
\usepackage{units}
\usepackage{url}
\usepackage{appendix}
\usepackage{footnote}
\usepackage{natbib}

\newcolumntype{L}[1]{>{\raggedright\let\newline\\\arraybackslash\hspace{0pt}}p{#1}}
\newcolumntype{C}[1]{>{\centering\let\newline\\\arraybackslash\hspace{0pt}}p{#1}}
\newcolumntype{R}[1]{>{\raggedleft\let\newline\\\arraybackslash\hspace{0pt}}p{#1}}

\setstcolor{red}
\definecolor{change}{RGB}{255,3,3}

\newcommand{\refsec}[1]{$\S$\ref{#1}}

\newcommand{\hmpc}{\mbox{$h^{-1}\textrm{Mpc}$} }
\newcommand{\kmsMpc}{\mbox{$\textrm{km}\ \textrm{s}^{-1}\ \textrm{Mpc}^{-1}$}}
\newcommand{\hmpcthreeinv}{\mbox{$h^{3} \textrm{Mpc}^{-3}$}}

\newcommand{\gr}{\ensuremath{(g-r)}\xspace}

\newcommand{\xirp}{$\xi(r_{p})$}
\newcommand{\rband}{\textit{r}-band}

\newcommand{\MD}[1]{{\texttt Mr#1-SDSS}\xspace}

\newcommand{\mgroup}{\ensuremath{M_\textrm{group}}\xspace}
\newcommand{\MPA}{MPA-JHU }
\newcommand{\mcf}{\ensuremath{\mathcal{M}(r_{p})}\xspace}
\newcommand{\mcft}{\texorpdfstring{\mcf}{MCF}}
\newcommand{\oneh}{{1-halo}\xspace}
\newcommand{\twoh}{{2-halo}\xspace}
\newcommand{\sersic}{{S\'ersic}\xspace}
\newcommand{\fracdiff}[1]{\Delta f_{\mathrm{#1}}}
\newcommand{\fracval}[4]{P\left(\textrm{#1}=\textrm{#2}\ |\ \textrm{#3}=\textrm{#4}\right)}
\newcommand{\fracres}[2]{\frac{ {\fracdiff{\mathrm{#1}}} - \overline{\fracdiff{\mathrm{#2}}} }{\sigma_{\mathrm{#2}}}}

\newcommand{\fracdiffm}[1]{$\Delta f_{\mathrm{#1}}$}
\newcommand{\fracvalm}[4]{$P\left(\textrm{#1}=\textrm{#2}\ |\ \textrm{#3}=\textrm{#4}\right)$}

\newcommand{\fgalnew}[4]{P\left(\textrm{#1}=\textrm{#2} | \textrm{#3}=\textrm{#4}\right)}
\newcommand{\fgalnewm}[6]{$\fgalnew$}
\newcommand{\sig}[1]{$#1\sigma$}
\newcommand{\rp}{$r_{p}$ }
\newcommand{\mcfeq}{\mathcal{M}(r_{p})}

\newcommand{\ssfr}{{sSFR}\xspace}
\newcommand{\sfr}{{SFR}\xspace}

\newcommand{\catsurl}{\url{https://github.com/vcalderon2009/SDSS_Conformity_Analysis}}

\defcitealias{Weinmann2006}{W06}
\defcitealias{Kauffmann2013a}{K13}
\defcitealias{Tinker2017a}{T17}
\defcitealias{Sin2017}{S17}
\defcitealias{Zu2016}{Z16}

\input{macros_dc.sty}

\definecolor{orange}{rgb}{1,0.5,0}
\let\oldmarginpar\marginpar
\renewcommand\marginpar[1]{\-\oldmarginpar[\raggedleft\footnotesize #1]%
{\raggedright\footnotesize #1}}

\usepackage[markup=underlined, final]{changes}
\definecolor{orange}{rgb}{1,0.5,0}
\definechangesauthor[color=red]{VC}
\usepackage{todonotes}
\setremarkmarkup{\todo[color=Changes@Color#1!20,size=\scriptsize]{#1: #2}}
\newcommand{\note}[2][]{}

\title{Small- and Large-Scale Galactic Conformity in SDSS DR7}
\author[Calderon et al.]{%
Victor F. Calderon$^{1}$\thanks{E-mail:  \mailto{victor.calderon@vanderbilt.edu}},
Andreas A. Berlind$^{1}$,
Manodeep Sinha$^{1, 2,3}$\vspace*{0.2em} \\
$^{1}$Department of Physics and Astronomy, Vanderbilt University, Nashville, TN 37235, USA\\
$^{2}$Centre for Astrophysics and Supercomputing, Swinburne University of Technology, Hawthorn, Victoria 3122, Australia\\
$^{3}$ARC Centre of Excellence for All Sky Astrophysics in 3 Dimensions (ASTRO 3D)
}

\date{Accepted XXX. Received YYY; in original form ZZZ}

\pubyear{2017}

\begin{document}
\setstcolor{red}
\label{firstpage}
\pagerange{\pageref{firstpage}--\pageref{lastpage}}
\maketitle

\begin{abstract}
Galactic conformity is the phenomenon whereby galaxy properties exhibit
excess correlations across distance than that expected if these properties
only depended on halo mass. We perform a comprehensive study of conformity
at low redshift using a galaxy group catalogue from the SDSS DR7
spectroscopic sample. We study correlations both between central galaxies
and their satellites (1-halo), and between central galaxies in separate haloes
(2-halo). We use the quenched fractions and the marked correlation function
(MCF), to probe for conformity in three galaxy properties, $\gr$ colour,
specific star formation rate (sSFR), and morphology. We assess the
statistical significance of conformity signals with a suite of mock galaxy
catalogues that have no built-in conformity, but contain the same
group-finding and mass assignment errors as the real data. In the case of
1-halo conformity, quenched fractions show strong signals at all group
masses. However, these signals are equally strong in mock catalogues,
indicating that the conformity signal is spurious and likely entirely
caused by group-finding systematics, calling into question previous claims
of 1-halo conformity detection. The MCF reveals a significant detection
of radial segregation within massive groups, but no evidence of conformity.
In the case of 2-halo conformity, quenched fractions show no significant
evidence of conformity in colour or \ssfr once compared with mock catalogues,
but a clear signal using morphology. In contrast, the MCF reveals a small,
yet highly significant signal for all three properties in low mass groups
and scales of $0.8-4\hmpc$, possibly representing the first robust detection
of 2-halo conformity.
\end{abstract}



\section{Introduction}
\label{sec:intro}

Characterizing the relation between the properties of galaxies and their
host dark matter (DM) haloes -- referred to as the ``galaxy-halo''
connection -- has emerged as a powerful tool to constrain theories of
galaxy formation with statistical measurements in galaxy surveys.
The phenomenon called ``\textit{galactic conformity}'' is a subtle feature
of this galaxy-halo connection, whereby galaxy properties are spatially
correlated even {\em at fixed halo mass}. Specifically, several studies
have claimed to detect a correlation between quenching properties
of galaxies, such as morphology, gas content, star formation rate, neutral
hydrogen content, and broad-band colour, and those of neighbouring galaxies
\citep{Weinmann2006,Ann2008,Ross2009,Kauffmann2010,Prescott2011,Wang2012,
Kauffmann2013a,Knobel2015,Hartley2015,Wang2015,Kawinwanichakij2016,Berti2017,
Zu2017}. This effect of ``galactic conformity'' exists over two distance
regimes, both between central and satellite galaxies within the same halo,
and between galaxies separated by several virial radii of their haloes.
We refer to these regimes as ``\oneh'' and ``\twoh'' conformity,
respectively \citep{Hearin2015b}. 2-halo conformity is closely linked to
``halo assembly bias'' or ``secondary bias''
\citep[e.g.,][]{Gao2005,Wechsler2006,Salcedo2017},
whereby the clustering of haloes depends on secondary properties, like age,
at fixed mass, and ``galaxy assembly bias''
\citep[e.g.,][]{Croton2007}, whereby galaxies inherit this clustering when
their observed properties correlate with these secondary halo properties.
Assembly bias provides a natural explanation for 2-halo conformity
\citep{Hearin2015b,Hearin2016}.

Conformity detections are notoriously difficult to make because it is hard 
to be confident that measurements are truly being made at fixed halo mass
and also to know whether a given galaxy pair lives in the same halo or not.
At the present time, there are several detection claims of \oneh conformity
at both low and high redshifts. These studies have looked at correlations 
between galaxy properties of the central galaxies and their respective 
satellite galaxies. Some have used isolation criteria to distinguish between 
centrals and satellites, while others have used group galaxy catalogues to 
do this. However, the impact of systematic errors on these results has not 
been quantified. In the \twoh regime, conformity has not yet been detected, 
as a couple recent works showed convincingly that past detections were 
entirely caused by selection biases. The current state of affairs for both 
\oneh and \twoh galactic conformity is still inconclusive and it is thus 
important to investigate this further.

The term ``galactic conformity'' was first coined by
\citet[][hereafter \citetalias{Weinmann2006}]{Weinmann2006} after
finding a correlation between the colours and star formation rates
(SFR) of central and satellite galaxies in common \citet{Yang2005}
galaxy groups of similar mass at low-redshifts, i.e. $z < 0.05$,
in SDSS \citep{York2000b} DR2 \citep{Abazajian2004a}. Specifically,
\citetalias{Weinmann2006} found that in galaxy groups of similar mass,
quenched satellite galaxies occur more frequently around quenched
central galaxies than around star-forming central galaxies.
Controlling for halo mass is of critical importance in conformity
studies because the SFRs of both centrals and satellites decrease with
halo mass, which can naturally induce a conformity-like signal.
\citetalias{Weinmann2006} attempted to control for halo mass by adopting
bins in total group luminosity.

Several subsequent studies also found correlations in SFR and other
properties between central and satellite galaxies, using different
methods for distinguishing between centrals and satellites and different
ways of controlling for mass. \citet{Ann2008} used isolation criteria,
rather than a group catalogue, to identify centrals and satellites in
SDSS DR5 \citep{McCarthy2007}. They found that early-type satellite
galaxies tend to reside in the vicinity of early-type central galaxies,
and argue that this conformity in morphology is likely due to hydrodynamic
and radiative influence of central galaxies on satellite galaxies, in
addition to tidal effects. They attempted to control for mass by
restricting their analysis to central galaxies in a limited range
of luminosity. \citet{Wang2012} also used isolation
criteria to study correlations between isolated bright primary galaxies
 in SDSS DR7 \citep{Abazajian2008} and nearby secondary galaxies
 (i.e., satellites) in SDSS DR8 \citep{Aihara2011}. They found
that the colour distribution of satellites is redder for red primaries
than for blue primaries of the same stellar mass. This is a similar
1-halo conformity trend in colour as found by \citetalias{Weinmann2006},
except that \citet{Wang2012} control for central galaxy stellar mass.
In addition, \citet{Wang2012} compared their results to the \citet{Guo2011}
semi-analytic model (SAM). They found that the SAM predicted a similar
conformity signal as the SDSS. However, when they re-analysed the SAM
controlling for halo mass instead of central galaxy stellar mass,
they found a substantially reduced signal. This implies that a large
portion of their observed SDSS conformity signal could be due to halo
mass differences between red and blue galaxies at fixed stellar mass.
\citet{Phillips2014,Phillips2015a} also used isolation criteria to
study the SFR of $\sim0.1L^{*}$ satellites around isolated $\sim L^{*}$
central galaxies in the local Universe using SDSS DR7. They found that
satellites of quiescent primaries are more than twice as likely to be
quenched than similar mass satellites of star forming primaries.
Unlike other studies, these authors control for the stellar mass of
satellites, rather than centrals. This might seem risky since
satellite galaxy stellar mass is not expected to correlate strongly
with halo mass. However, the authors compare the velocity distributions
of satellites around star forming and quiescent primaries and they
conclude that the difference in halo mass between the two samples is
not large enough to account for the conformity signal they observe.
Finally, \citet{Knobel2015} used the \citet{Yang2012} group catalogue in
SDSS DR7 to study the degree of central-satellite conformity, controlling
for several combinations of properties, including total group stellar mass.
They confirmed that satellites of quenched central galaxies are more
likely to be quenched than those of active central galaxies.

\citet{Ross2009} found evidence of 1-halo conformity using a
completely different approach. They used a Halo Occupation Distribution
\citep[HOD;][]{Berlind2002} model to fit the clustering of photometric
samples in SDSS DR5. They found that they could only simultaneously
match the clustering of all, early- and late-type galaxies with a
model that segregates early- and late-type galaxies into separate
haloes as much as possible. This is similar in spirit to the previous
work of \citet{Zehavi2004a} who modelled the cross-correlation function
between red and blue galaxies in SDSS DR2, though that study concluded
that red and blue galaxies are well mixed within their haloes.
\citet{Zehavi2010} revisited this issue using SDSS DR7 and found
significant evidence of colour segregation into different
haloes, but the degree of segregation was much less than that
found by \citet{Ross2009}.

There have also been studies that have claimed a detection of
1-halo conformity at higher redshift. \citet{Hartley2015} used
isolation criteria in the UKIDSS \citep{Lawrence2007} Ultra Deep
Survey DR8, to explore the redshift evolution of the correlation
between the \sfr of central galaxies and satellite galaxies at
intermediate to high redshifts ($0.4<z<1.9$). They confirmed that
passive satellites tend to be preferentially located around
passive central galaxies, and showed that the trend persists to at
least $z\sim2$ without any significant evolution. \citet{Kawinwanichakij2016}
carried out a similar analysis and identified central and
satellite galaxies in the range of $0.3 < z < 2.5$ by combining
imaging from three deep near-infrared-selected surveys ZFOURGE/CANDELS,
UDS, and UltraVISTA \citep{McCracken2012} and deriving accurate
photometric redshifts. They found that, at similar central
stellar mass, satellites of quiescent central galaxies are more
likely to be quenched compared to satellites of star-forming
central galaxies. This conformity signal is only significant at
$0.6<z<1.6$, and becomes weaker at both lower and higher redshifts.
\citet{Kawinwanichakij2016} argue that their detection is unlikely
to arise from any difference in halo mass between star-forming
and quiescent centrals. To check this they allowed for star-forming
centrals to have a stellar mass of up to 0.2 dex higher than quiescent
centrals and found that, though the conformity signal weakens,
it does not vanish. Most recently, \citet{Berti2017} used isolation
criteria in the spectroscopic PRIMUS Survey \citep{Coil2011,Cool2013}
to look for conformity at $0.2<z<1.0$. After matching the stellar mass
and redshift distributions of star-forming and quenched centrals,
\citet{Berti2017} claimed a \sig{3} detection of a $\sim5\%$ excess
of star-forming neighbours around star-forming central galaxies
on scales of 0-1 Mpc. This conformity signal is substantially
weaker than the \citetalias{Weinmann2006} signal observed in
SDSS at $z \lesssim 0.05$. \citet{Berti2017} also reported on a
2-halo conformity detection, albeit with weaker statistical significance.

In the 2-halo regime,
\citet[][hereafter \citetalias{Kauffmann2013a}]{Kauffmann2013a}
claimed a detection of conformity using a volume-limited
sample of galaxies with redshifts $z<0.03$ from the SDSS DR7.
They adopted isolation criteria to identify central galaxies and
studied the median specific SFR of neighbouring galaxies as a
function of different properties of the centrals.
\citetalias{Kauffmann2013a} found that the \sfr of neighbours
correlates with that of centrals, even up to 4 Mpc, a distance
that is well outside the virial radius of the primary galaxy's
halo. This 2-halo conformity signal is present for low stellar
mass galaxies, with massive galaxies only exhibiting a
1-halo conformity signal. The \citetalias{Kauffmann2013a} result
was intriguing and motivated a number of theoretical studies to
explain it. However, a pair of recent studies have 
shown convincingly that the result in
\citetalias{Kauffmann2013a} is entirely due to selection bias.
\citet[][hereafter \citetalias{Tinker2017a}]
{Tinker2017a} reproduced the result of \citetalias{Kauffmann2013a}
and then used a group finding algorithm and a mock catalogue to
show that the majority of the 2-halo conformity signal comes from a
subset of satellite galaxies that were mis-identified as primaries
in the galaxy sample. After removing this small fraction of satellite
galaxies, \citetalias{Tinker2017a} detect no statistically significant
2-halo conformity in galaxy star formation rates.
\citet[][hereafter \citetalias{Sin2017}]{Sin2017} carried out
a similar analysis, and argued that the isolation criteria in
\citetalias{Kauffmann2013a} could potentially include low-mass
central galaxies in the vicinity of massive systems, and that
the large-scale conformity signal is likely a short-range effect coming
from massive haloes. In addition to the misclassification of
satellite galaxies as central galaxies in the isolation criteria,
\citetalias{Sin2017} argued that a weighting in favour of central
galaxies in very high-density regions, and the use of medians to
characterize the bimodal distribution of \ssfr could potentially amplify
the large-scale conformity signal seen in \citetalias{Kauffmann2013a}.

\citet{Zu2017} came to a similar conclusion about the lack of
2-halo conformity by finding that conformity measurements in
SDSS DR7 are consistent with predictions from the {\texttt iHOD}
\textit{halo-quenching} model ~\citep{Zu2015,Zu2016}, in which
galaxy colours depend \textit{only} on halo mass. This suggests
that all conformity signals are simply due to the combination of
the environmental dependence of the halo mass function combined
with the strong correlation between galaxy colours and halo mass.
In other words, no galaxy assembly bias or other environmental
quenching mechanisms are required to explain 2-halo conformity signals.

On the theoretical side, there have been several studies looking
at both 1-halo and 2-halo conformity. \citet{Paranjape2015} called
into question the conformity signal measured by
\citetalias{Kauffmann2013a} at a projected distance of
$\lesssim 4$ Mpc by generating mock catalogues with varying
levels of built-in galactic conformity, and comparing these to
SDSS galaxies in the \citet{Yang2007} galaxy group catalogue.
They argued against the \citetalias{Kauffmann2013a} result being
evidence of galaxy and halo assembly bias. \citet{Paranjape2015}
also argued that only at very large separations, ($\gtrsim 8$ Mpc),
does 2-halo conformity, driven by the assembly bias of small haloes,
manifest distinctly. They suggest that the observed conformity at
$\lesssim 4$ Mpc is simply due to central galaxies of similar
stellar mass residing in haloes of different masses. Other
papers have tried to explore the origin of galactic conformity.
\citet{Hearin2016} studied the correlation between the mass
accretion rates of nearby haloes as a potential physical origin for
2-halo galactic conformity. They found that pairs of host haloes
have correlated assembly histories, despite being separated from
each other by distances greater than thirty virial radii at
the present day. They presented halo accretion conformity as
a plausible mechanism driving 2-halo conformity in SFR.
Moreover, they argued that galactic conformity is related to
large-scale tidal fields, and predicted that 2-halo conformity
should generically weaken at higher redshift and vanish to
undetectable levels by $z\sim 1$. In this context, the 2-halo
galactic conformity signal in \citet{Berti2017} is consistent with
the \citet{Hearin2016} prediction and \citet{Berti2017} state that their
detection of galactic conformity is thus likely indicative of assembly
bias and arises from large-scale tidal fields. Additionally,
\citet{Bray2016} investigated the role of assembly bias in producing
galactic conformity in the Illustris \citep{Vogelsberger2014}
simulation, and argued to have found 2-halo conformity in the
red fraction of galaxies. They found that, at fixed stellar mass,
the red fraction of galaxies around redder neighbour galaxies is
higher than it is around bluer galaxies and this effect persists
out to distances of 10 Mpc. They concluded by saying that the
predicted amplitude of the conformity signal depends on the
projection effects, stacking techniques, and the criteria used for
selecting central galaxies. \citet{Lacerna2017} used three semi-
analytic models to study the correlations between \ssfr
of central galaxies and neighbour galaxies out to scales of several Mpc.
They predicted a strong 1-halo galactic conformity signal when the
selection of primary galaxies was based on an isolation criterion
in real space, and claimed a significant 2-halo conformity signal
as far as $\sim5$ Mpc. However, the overall signal of galactic
conformity decreased when satellites that had been misclassified as
central galaxies were removed in the selection of primary galaxies.
The authors concluded that the SAMs used in the analysis do not show
galactic conformity for the 2-halo regime.

Galactic conformity remains a debated topic, and it is unclear if
all previous detection claims are valid. The work of
\citet{Campbell2015d} exposed the dangers of using group catalogues
to study 1-halo conformity. They showed that group finders do
a good job at recovering galactic conformity, but they also tend
to introduce weak conformity when none is present in the data.
This calls into question previous claims, such as the one by
\citetalias{Weinmann2006}. More recently, \citetalias{Tinker2017a}
and \citetalias{Sin2017} challenged the measurement of 2-halo
conformity made by \citetalias{Kauffmann2013a} by showing that
their isolation criteria were not sufficiently robust.
These conflicting results open the door for improvements in the
measurements of 1-halo and 2-halo conformity. In this paper we
investigate both regimes using a galaxy group catalogue from the
SDSS DR7. Our analysis contains three main improvements over
previous works. First, we study three observed properties of
galaxies: \gr colour, \ssfr, and \sersic index. Second, we use a
new statistic, the marked correlation function, \mcf, in addition
to the previously used quenched fractions. \mcf is ideally
suited for conformity studies and is a more sensitive probe
of weak conformity signals. Third, we use a suite of 100 mock
galaxy catalogues to quantify the statistical significance of
our results. The mock catalogues do not have any built-in conformity,
but they are affected by the same systematic errors as the
SDSS data. By comparing our SDSS measurements to the distribution
of mock measurements, we can quantify the probability that
whatever signal we detect could have arisen from a model with no conformity.

This paper is organized as follows. In \refsec{sec:data_meas},
we describe the observational and simulated data used in this work,
as well as the main analysis methods. In \refsec{sec:Gal_Conf},
we present a detailed examination of galactic conformity,
distinguishing between 1-halo (\refsec{sub:one_halo_conf}) and 2-halo
(\refsec{sub:two_halo_conf}). We summarize our results and discuss their
implications in \refsec{sec:summary_discussion}. The Python codes
and the catalogues used in this project will be made publicly
available on Github \footnote{\catsurl} upon publication of this paper.


\section{Data and Methods}
\label{sec:data_meas}

In this section, we present the datasets used throughout this analysis, and
introduce the main statistical methods that we use to search for conformity
signals. In \refsec{subsec:NYU_DR7} we briefly describe the SDSS galaxy sample
that we use, along with the parameters that are included in this catalogue. In
\refsec{sub:group_finding_mass} we summarize how we identify galaxy groups and
estimate their masses. In \refsec{subsec:Mock_Catls} we describe in detail the
mock catalogues that we use throughout the paper. Finally, we describe the two
main statistical methods for probing conformity in
\refsec{sub:quenched_frac_method} and \refsec{sub:marked_corr}.

\subsection{SDSS Galaxy Sample}
\label{subsec:NYU_DR7}

For this analysis, we use data from the Sloan Digital Sky Survey.
SDSS collected its data with a dedicated 2.5-meter telescope
\citep{Gunn2006}, camera \citep{Gunn1998}, filters \citep{Doi2010},
and spectrograph \citep{Smee2012}. We construct our galaxy
sample from the {\texttt{large-scale structure}} sample of the
NYU Value-Added Galaxy Catalogue \citep[NYU-VAGC;][]{Blanton2005},
based on the spectroscopic sample in Data Release 7
\citep[SDSS DR7;][]{Abazajian2008}. The main spectroscopic galaxy
sample is approximately complete down to an apparent \rband\
Petrosian magnitude limit of $m_{r} = -17.77$. However, we have
cut our sample back to $m_{r} = -17.6$ so it is complete down to
that magnitude limit across the sky. Galaxy absolute magnitudes are
\textit{k}-corrected \citep{Blanton2003a} to rest-frame magnitudes
at redshift $z=0.1$.

We construct a volume-limited galaxy sample that contains all
galaxies brighter than $M_r = -19$, and we refer to this sample
as \MD{19}. The redshift limits of the sample are
$z_\mathrm{min}=0.02$ and $z_\mathrm{max}=0.067$ and it contains 90,893
galaxies with a number density of $n_\mathrm{gal}=0.01503$\hmpcthreeinv.
The sample includes the right ascension, declination, redshift,
\sersic index, and \gr colour for each galaxy.

To each galaxy, we assign a star formation rate (\sfr) using the
MPA Value-Added Catalogue DR7
\footnote{\url{http://www.mpa-garching.mpg.de/SDSS/DR7}}.
This catalogue includes, among many other parameters, stellar
masses based on fits to the photometry using \cite{Kauffmann2003a}
and \cite{Salim2007}, and star formation rates based on
\cite{Brinchmann2004}. We cross-match the galaxies of
the NYU-VAGC catalogue to those in the \MPA catalogue using their
MJD, plate ID, and fibre ID. A total of 5.85\% of galaxies in the
sample did not have corresponding values of \sfr and we remove them
from the sample. This leaves a sample of 85,578 galaxies.
For each of these galaxies, we divide its \sfr by its stellar mass to
get a specific star formation rate \ssfr.

\ssfr and \gr colour are highly correlated galaxy properties with the main 
difference coming from dust attenuation that moves some intrinsically star
forming galaxies onto the red sequence. However, we have chosen to use both
galaxy properties in this analysis in order to facilitate the comparison of 
our work to previous claims of conformity detection.

\subsection{Group Finding Algorithm and Mass Assignment}
\label{sub:group_finding_mass}

We identify galaxy groups using the \cite{Berlind2006} group-finding
algorithm. This is a Friends-of-Friends \citep[FoF;][]{Huchra1982a} algorithm
that links galaxies recursively to other galaxies that are within a
cylindrical linking volume. The projected and line-of-sight linking lengths
are $b_{\perp} = 0.14$ and $b_{\parallel} = 0.75$ in units of the mean
inter-galaxy separation. This choice of linking lengths was optimized by
\cite{Berlind2006} to identify galaxy systems that live within the same
dark matter halo. In each group, we define the most luminous galaxy
(in the \rband) to be the `central' galaxy. The rest of the galaxies are
defined as `satellite' galaxies.

We estimate the total masses of the groups via \textit{abundance matching},
using total group luminosity as a proxy for mass. Specifically, we assume
that the total group \rband\ luminosity $L_{\textrm{group}}$ increases
monotonically with halo mass $M_{\textrm{h}}$, and we assign masses to groups
by matching the cumulative space densities of groups and haloes:
\begin{align}
n_{\textrm{group}} (> L_{\textrm{group}}) &=
n_{\textrm{halo}}(> M_{\textrm{h}}).
\end{align}
To calculate the space densities of haloes, we adopt the \cite{Warren2006}
halo mass function assuming a cosmological model with
$\Omega_{m} = 1 - \Omega_{\Lambda} = 0.25$, $\Omega_{b} = 0.04$,
$h \equiv H_{0}/$ (100 \kmsMpc) = 0.7, $\sigma_{8} = 0.8$, and $n_{s} = 1.0$.
We refer to these abundance matched masses as \textit{group masses}, \mgroup.

\subsection{Mock Galaxy Catalogues}
\label{subsec:Mock_Catls}

To quantify the statistical significance of any conformity signal that we
measure using our SDSS groups, it is necessary to compare to a null model
(i.e., no intrinsic conformity) that incorporates the same systematic errors
as our measurements and also includes robust error distributions.
For this purpose, we construct a suite of 100 realistic mock catalogues that
are based on the \textit{Large Suite of Dark Matter Simulations} (LasDamas)
project\footnote{\url{http://lss.phy.vanderbilt.edu/lasdamas/}}
\citep{McBride2009}.

We start with a set of 50 cosmological N-body simulations that trace the
evolution of dark matter in the Universe and have sufficient volume and mass
resolution to properly model the \MD{19} sample. These simulations assumed
the same cosmological model described at the end of
\refsec{sub:group_finding_mass}. Dark matter haloes were identified with a
FoF algorithm \citep{Davis1985a} using a linking length of 0.2 times the
mean inter-particle separation. We used an HOD model to populate the DM
haloes with central and satellite galaxies, whose numbers as a function of
halo mass were chosen to reproduce the number density, 
$n_\mathrm{gal}$, and the projected 2-point correlation function,
$w_{p}\left(r_p\right)$, of the \MD{19} sample. Each central galaxy was
placed at the minimum of the halo gravitational potential and was
assigned the mean velocity of the halo. Satellite galaxies were assigned
the positions and velocities of randomly chosen dark matter particles within
the halo. Within each simulation box, we applied redshift space distortions
and then we carved out two independent volumes that precisely mimic the
geometry of our \MD{19} sample. This procedure yields 100 independent mock
catalogues from the 50 simulation boxes.

To assign a luminosity to each mock galaxy, we adopt the formalism of the
\textit{conditional luminosity function} \citep[CLF;][]{Yang2003,
VanDenBosch2003} that specifies functional forms for the luminosity
distributions of central and satellite galaxies as a function of halo mass.
Specifically, we use the \citet{Cacciato2009} version of the CLF, but modified
slightly to match our adopted cosmological model (Van den Bosch, private
communication). We then abundance match the luminosities obtained from the
CLF to the \rband\ absolute magnitudes in \MD{19}. As a result, our mock
catalogues have the same exact luminosity function as the SDSS data.

We assign specific star formation rates, \gr colours and
\sersic indices to mock galaxies by first adopting the formalism presented in
\citet[][hereafter \citetalias{Zu2016}]{Zu2016}, and then sampling from the
original distributions of \ssfr, \gr colour and \sersic indices
of \MD{19}. Specifically, we adopt the \textit{`halo'} quenching model from
\citetalias{Zu2016}, which assumes that halo mass is the sole driver of galaxy
quenching. According to that model, the red/quenched fraction of central and
satellite galaxies is given by
\begin{align}\label{eq:zu_cen}
    f_{\rm{cen}}^{\rm{red}}(M_\textrm{h}) &= 1-\exp\left[-(M_\textrm{h} / M_\textrm{h}^\textrm{qc})^{\mu^\textrm{c}}\right]
\end{align}
and
\begin{align}\label{eq:zu_sat}
    f_{\rm{sat}}^{\rm{red}}(M_\textrm{h}) &=1-\exp\left[-(M_\textrm{h} / M_\textrm{h}^\textrm{qs})^{\mu^\textrm{s}}\right] ,
\end{align}
where $M_\textrm{h}^\textrm{qc}$, $M_\textrm{h}^\textrm{qs}$, $\mu^\textrm{c}$,
and $\mu^\textrm{s}$ are parameters of the model that \citetalias{Zu2016}
fit to the observed clustering and galaxy-galaxy lensing measurements of
red and blue galaxies in the SDSS. We assign each of our mock galaxies a
probability of being quenched from equations~(\ref{eq:zu_cen})
and~(\ref{eq:zu_sat}) and we randomly designate it as `active' or `passive'
consistent with that probability (e.g., if $f_{\rm{sat}}^{\rm{red}}=0.8$
for a particular mock satellite galaxy, we give it an 80\% chance of being
labelled `passive'). 
To assign realistic values of \ssfr, \gr colour, and \sersic index to 
mock galaxies, we divide the observed distributions of these properties
of \MD{19} into `active' and `passive' distributions by making cuts 
at $\log_{10} \textrm{\ssfr} = -11$, $(g-r)_{cut} = 0.75$ and $n_{cut} = 3$
for \ssfr, \gr colour, and \sersic index, respectively. For example, to assign
\ssfr values to mock galaxies, we do the following. For each mock galaxy,
we randomly draw a \ssfr value from the active or passive distribution,
depending on the designation that the mock galaxy has received. Moreover,
we do this in a way that preserves the joint \ssfr-luminosity distribution.
For example, if a mock galaxy has been labelled `active', we randomly
select a real active galaxy from \MD{19} that has a similar luminosity as the
mock galaxy, and we assign its sSFR to the mock galaxy. As a result of this
procedure, the final joint sSFR-luminosity distribution of mock galaxies
closely resembles the one for \MD{19}. However, the model contains
{\em no intrinsic} 1-halo or 2-halo conformity because the galaxy
sSFR values {\em only} depend on halo mass.
We apply this same procedure to assign \gr colours and \sersic
indices to each mock galaxy in order to preserve the joint distributions
of these galaxy properties with luminosity as seen the \MD{19} sample.

After constructing our 100 mock catalogues, we run the group-finding
algorithm on each one to produce a corresponding group catalogue. We then
label each mock galaxy as `central' or `satellite' and estimate total group
masses by following the same methodology as in
\refsec{sub:group_finding_mass}. The end result is a set of mock catalogues
that do not have built-in galactic conformity in \ssfr, \gr colour, or
\sersic index, but suffer from the same kinds of systematics as the SDSS
data, i.e. group-finding errors that lead to central-satellite
misclassification and errors in the estimated group masses.

\subsection{Quenched Fraction Difference \texorpdfstring{\fracdiffm{q}}{fracdiff}}
\label{sub:quenched_frac_method}

Previous studies of conformity have mostly focused on measuring the fractions
of quenched neighbour galaxies around active and passive primary galaxies,
either as a function of group mass or as a function of distance (e.g.,
\citetalias{Weinmann2006}, \citetalias{Kauffmann2013a}). Following these
studies, we also consider quenched fractions of neighbour galaxies, focusing
on the \textit{difference} between the fraction for passive primaries and that
for active primaries. Moreover, we use three different galaxy properties to
search for conformity: \gr colour, \ssfr, and \sersic index. The cuts we use
to designate galaxies as red, passive or early type are $(g-r) > 0.75$,
$\log\ \textrm{sSFR} < -11$, and $n > 3$, respectively. These are the same
cuts we discuss in \refsec{subsec:Mock_Catls}.

To explain this better, let us consider the specific case of probing 1-halo
conformity in galaxy colour. We measure the fraction of red satellite galaxies
around red centrals, \fracvalm{sat}{red}{cen}{red}, and the fraction of red
satellite galaxies around blue centrals, \fracvalm{sat}{red}{cen}{blue}. We
then determine the \textit{difference} between these two fractions, which we
refer to as \fracdiffm{red}. A conformity signal is then the case of $\left|
\fracdiff{red}\right|  > 0$. We define similar quantities using \ssfr and
morphology. The three quenched fraction differences that we measure are thus
\begin{align}
\fracdiff{red} & =  \fracval{sat}{red}{cen}{red} \label{eq:fgalrat_col}\\ & \qquad - \fracval{sat}{red}{cen}{blue}\nonumber\\[3pt]
\fracdiff{passive}& = \fracval{sat}{passive}{cen}{passive} \label{eq:fgalrat_ssfr}\\ & \qquad - \fracval{sat}{passive}{cen}{active}\nonumber\\[3pt]
\fracdiff{early} & = \fracval{sat}{early}{cen}{early} \label{eq:fgalrat_morph}\\ & \qquad - \fracval{sat}{early}{cen}{late}\nonumber
\end{align}
Finally, as a way to control for halo mass, we measure these fractions in
bins of \mgroup. In the mock catalogues, we follow the same procedure
to calculate $\fracdiff{red}$, $\fracdiff{passive}$ and $\fracdiff{early}$
. For convenience, we refer to all three of these
quantities as ``quenched'' fraction differences, $\fracdiff{q}$, recognizing
that \sersic index is a measure of galaxy morphology and not star formation
activity.


In the case of 2-halo conformity, we use the same formalism of
equations~(\ref{eq:fgalrat_col})$-$(\ref{eq:fgalrat_morph}), with the
difference that we only consider pairs of central galaxies with line-of-sight
separations of $\pi_\mathrm{max}<20\mpch$ and we calculate the fractions in
bins of projected separation within each \mgroup bin. For each
central-central galaxy, we designate one to be the primary and the other to
be the secondary and we calculate the difference between the quenched
fractions of secondary galaxies that are associated with quenched primaries
and those that are associated with active primaries. Each galaxy pair
contributes twice to the calculation of $\fracdiff{q}$ because both galaxies
get a turn at being considered the primary galaxy. For example, suppose there
is a pair of galaxies, one red
and one blue, that are both centrals in groups of similar mass. When the
blue galaxy is the primary, the pair will contribute positively to the
fraction \fracvalm{secondary}{red}{primary}{blue}. On the other hand, when
the red galaxy is the primary, the pair will contribute negatively to the
fraction \fracvalm{secondary}{red}{primary}{red}. Therefore, red-red and
blue-blue pairs act to increase $\fracdiff{red}$, while red-blue pairs do
the opposite. $\fracdiff{red}$ essentially measures the excess number of
similar pairs (i.e., red-red or blue-blue) over what one would expect if
the population of red and blue galaxies were randomly mixed. The value of
$\fracdiff{q}$ ranges from $+1$ where all pairs are similar, to $-1$ where
pairs are as different as possible.

\subsection{Marked Correlation Function  \texorpdfstring{\mcf}{MCF}}
\label{sub:marked_corr}

Galactic conformity is essentially a correlation between the properties of
galaxies across distance. In the case of 1-halo conformity, we care about
the correlation between properties of central galaxies and satellites within
the same halo. In the case of 2-halo conformity, we look for a correlation
between properties of central galaxies in separate haloes. The
``\textit{marked correlation function}'' is an ideal tool for quantifying
correlations across scale and it has been used successfully to probe the
environmental dependence of galaxy properties
\citep{Beisbart2000,Sheth2005a,Skibba2006,Martinez2010}.

The marked statistic \mcf provides a measure of the clustering of galaxy
properties, or ``marks''. In this paper, we analyse the marked statistics
for \gr colour, specific star formation rate (\ssfr), and \sersic index $n$
in bins of group mass \mgroup. We adopt the formalism presented in
\citet{Sheth2005a} and \citet{Skibba2006} for defining \mcf
\begin{align}
\mcfeq &= \frac{1 + W(r_{p})}{1 + \xi(r_{p})} \equiv \frac{WW}{DD}
\label{eq:mcf}
\end{align}
where \xirp\ is the usual two-point correlation function with pairs summed
in bins of projected separation $r_{p}$, and $W (r_{p})$ is the same except
that galaxy pairs are weighted by the product of their marks. The estimator
used in equation~(\ref{eq:mcf}) can also be written as $WW/DD$, where $DD$
is the raw number of galaxy pairs separated by $r_{p}$ and $WW$ is the
weighted number of pairs. Defining the statistic as a ratio in this way is
advantageous because, unlike the correlation function, it can be estimated
without explicitly constructing a random catalogue, but, like the correlation
function, it accounts for edge effects so one does not need to worry about
the geometry of the survey \citep{Sheth2005a}.

The marked statistic is essentially a measurement of the correlation
coefficient between the marks of galaxies, as a function of projected
separation. Though it is similar in spirit and goal to the quenched fraction
difference statistic described in the previous section, the marked correlation
function contains more information because it uses the full values of galaxy
properties (e.g., colour) instead of just a binary classification (e.g., red
or blue). There is thus reason to hope that \mcf is a more sensitive probe
of galactic conformity than the usual quenched fractions.


\section{Galactic Conformity Results}
\label{sec:Gal_Conf}

In this section, we present the results of the galactic conformity analysis
of SDSS DR7. In \refsec{sub:one_halo_conf}, we investigate 1-halo conformity
by looking at both quenched fraction differences, $\fracdiff{q}$, as a
function of group mass (\refsec{subsub:one_halo_fracs}), and the mark
correlation function, \mcf, as a function of projected separation
(\refsec{subsub:one_halo_mcf}). In \refsec{sub:two_halo_conf}, we investigate
2-halo conformity, also using $\fracdiff{q}$
(\refsec{subsub:two_halo_conf_quench_frac}) and \mcf
(\refsec{subsub:two_halo_conf_mcf}).

\subsection{1-halo Conformity}
\label{sub:one_halo_conf}

\subsubsection{Quenched Fractions and 1-halo Conformity}
\label{subsub:one_halo_fracs}

We first study 1-halo conformity using the quenched fraction difference
statistic defined in \S~\ref{sub:quenched_frac_method} as a function of
group mass. This is very similar to the original method that
\citetalias{Weinmann2006} used to detect 1-halo conformity. Specifically, we
create six \mgroup bins of width 0.4 dex in the range log\mgroup : 11.6--14.0.
Within each bin of group mass, we make a list of all satellite galaxies
that are in groups with a red central and a second list of all satellites
in groups with a blue central. We then calculate the red fraction of
satellites in each list and take the difference $\fracdiff{red}$. We repeat
this process using \ssfr and \sersic index to calculate $\fracdiff{passive}$
and $\fracdiff{early}$. When using these quenched fraction differences, a
conformity signal corresponds to values that are not zero, i.e.,
$\left| \fracdiff{q}\right|  > 0$.

\begin{figure*}
    \centering
    \begin{centering}
    \includegraphics[width=0.95 \textwidth]{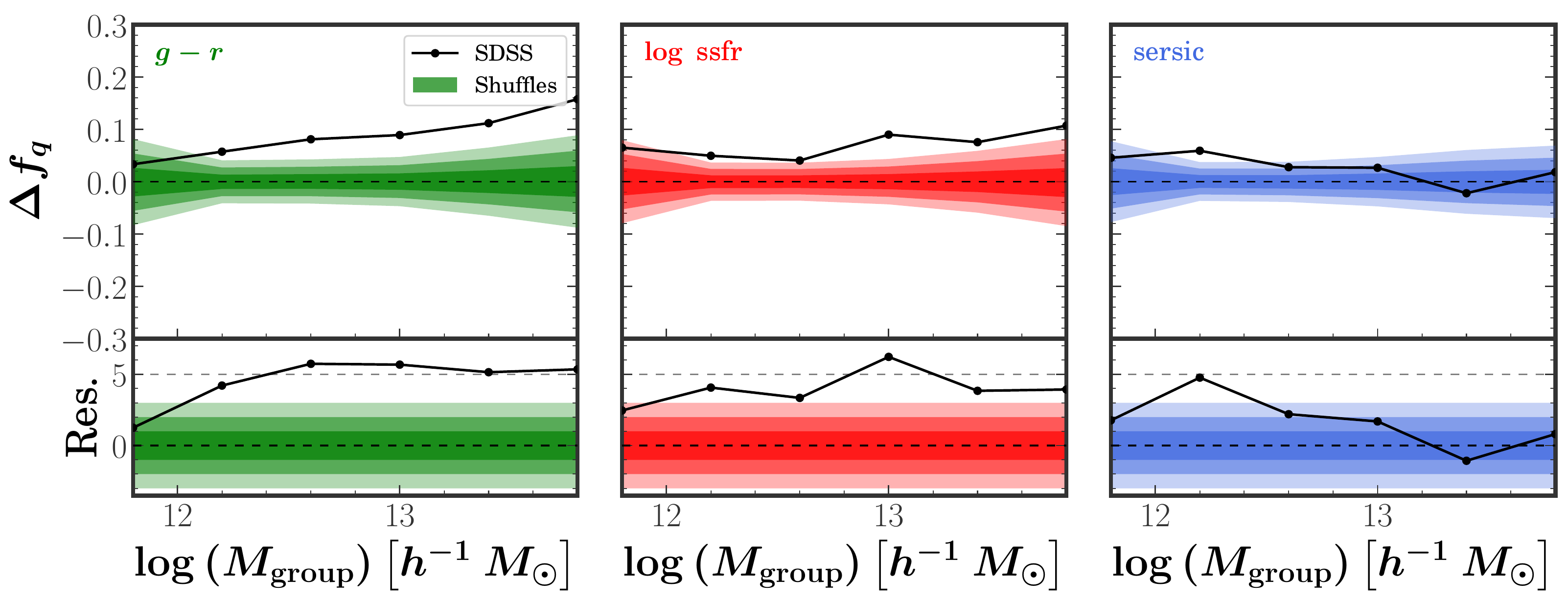}
    \end{centering}
    \caption{
    Difference of fractions, \fracdiffm{}, of red (left),
    passive (centre), and early-type (right) satellites
    as function of estimated group mass, \mgroup, where
    the difference is measured between groups with red and
    blue, passive and active, early-type and late-type
    central galaxies, as measured in the \MD{19} sample.
    \textit{Top panels}: The solid black lines correspond
    to the \fracdiffm{} of each galaxy property. The shaded
    contours show the \sig{1}, \sig{2}, and \sig{3} ranges
    of \fracdiffm{} calculated from many realizations in
    which the values of the galaxy properties are randomly
    shuffled, thus erasing any trace of 1-halo galactic
    conformity. \textit{Bottom panels}: Normalised residuals
    of \fracdiffm{} of each galaxy property with respect to
    the shuffled realizations. The solid black lines show
    the difference between \fracdiffm{} and the mean of the
    shuffles, divided by the standard deviation of
    \fracdiffm{} for the shuffles. The shaded contours show
    the \sig{1}, \sig{2}, and \sig{3} ranges of the shuffled
    scenario in this normalised space.
    }
    \label{Fig:one_halo_fractions_data}
\end{figure*}

To determine the statistical significance of any conformity signal, we use a
random shuffling method to eliminate any intrinsic conformity or correlation
in the sample at the group level. Specifically, we shuffle the properties
(colour, \ssfr, \sersic index) of all central and satellite galaxies within
each group mass bin. Each central galaxy swaps properties with a randomly
selected central galaxy from a different group of similar mass, and each
satellite swaps properties with a randomly selected satellite from a group
of similar mass. This procedure preserves the distributions of central and
satellite properties as a function of group mass, but it explicitly erases
any correlation between the properties of centrals and their satellites
within any single group. The shuffling thus completely erases any 1-halo
conformity signal that may exist in the data. We repeat this shuffling
process a total of 1000 times (using different random seeds) and we
re-measure the quenched fraction differences each time. The 
resulting distribution of $\fracdiff{q,shuffle}$ values thus allows us to
quantify the probability that any measured conformity signal could be a
statistical fluke. We find that the distribution of shuffle values is
consistent with being Gaussian and so we use the standard deviation of the
shuffled values to calculate the \sig{1}, \sig{2}, and \sig{3} ranges
of the distribution of $\fracdiff{q,shuffle}$. We adopt the \sig{3} level
as our detection threshold.

For each measurement of $\fracdiff{q}$ on the un-shuffled data, we calculate the
residual with respect to the shuffled data as
\begin{align}
\mathrm{Res} = \fracres{q}{q,shuffle} \label{eq:fgalres_eq}
\end{align}
where $\overline{\fracdiff{q,shuffle}}$ is the mean of the 1000 shuffles and
$\sigma_{\mathrm{q,shuffle}}$ is their standard deviation.

Figure~\ref{Fig:one_halo_fractions_data} presents our main results of probing
1-halo conformity using quenched fraction differences. The black lines in the
top three panels show the \fracdiffm{q} for \gr colour, \ssfr, and \sersic
index, as measured in the \MD{19} sample. The shaded contours show the
\sig{1}, \sig{2}, and \sig{3} ranges of \fracdiffm{shuffle} for the
\textit{shuffle} cases of each galaxy property. The bottom panels show the
residuals of each galaxy property with respect to the shuffles, as defined in
equation~(\ref{eq:fgalres_eq}). Figure~\ref{Fig:one_halo_fractions_data}
shows prominent conformity signals in the quenched fraction differences
for \gr colour and \ssfr at large group masses, while for morphology the
signal only appears at low group mass. Specifically,
the conformity signal in colour rises with mass from \fracdiffm{red}=0.06 to
0.14 and is at the $4-6\sigma$ level of statistical significance for masses
above $10^{12}\msunh$. In the case of \ssfr, the signal is lower, rising from
\fracdiffm{passive}=0.05 to 0.1 and is at the $3-4\sigma$ level, except for a
$6\sigma$ peak at $\sim 10^{13}\msunh$. Finally, in the case of S\'ersic
index, the signal is only significant for groups of mass
$\sim 10^{12.2}\msunh$, where \fracdiffm{early}=0.07 and has a statistical
significance of $5-6\sigma$. These results are in agreement with the results
shown in \citetalias{Weinmann2006}, who also find a significant difference
in the red fraction at high masses and a slightly weaker signal when
using \ssfr.

We have found statistically significant correlations between the properties of
central and satellite galaxies within groups in \MD{19} by comparing to the
distribution of shuffled measurements, where any correlations between the
properties of centrals and satellites have been erased. However, this does not
mean that we have detected 1-halo galactic conformity, which is a correlation
at fixed {\it halo} mass. Grouping errors that cause misidentification of
centrals and satellites as well as errors in the estimated group mass
\mgroup could be responsible for inducing a conformity-like signal
\citep{Campbell2015d}. To test this, we need to compare our measurements
to mock catalogues that contain no built-in conformity, but are analysed
in the same way as the SDSS data.

We apply the same procedure described above to the set of mock catalogues
described in \refsec{subsec:Mock_Catls}. The goal is to determine if the
signal revealed in Figure~\ref{Fig:one_halo_fractions_data}
remains statistically significant when compared to the distribution of
\fracdiffm{q,mock} measurements from the 100 mock catalogues with no
conformity built-in. We find that the distribution of 100 values of
\fracdiffm{q,mock} is approximately Gaussian and so we use their standard
deviation to estimate the $1\sigma$, $2\sigma$, and $3\sigma$ ranges of the
distribution. As we did previously for the shuffles, for each measurement of
$\fracdiff{q}$ on the SDSS data, we calculate the residual with respect to
the mocks as
\begin{align}
\mathrm{Res} = \fracres{q}{q,mock} \label{eq:fgalresmock_eq}
\end{align}
where $\overline{\fracdiff{q,mock}}$ is the mean of the 100 mocks and
$\sigma_{\mathrm{q,mock}}$ is their standard deviation.

\begin{figure*}
    \begin{centering}
    \includegraphics[width=0.95 \textwidth]{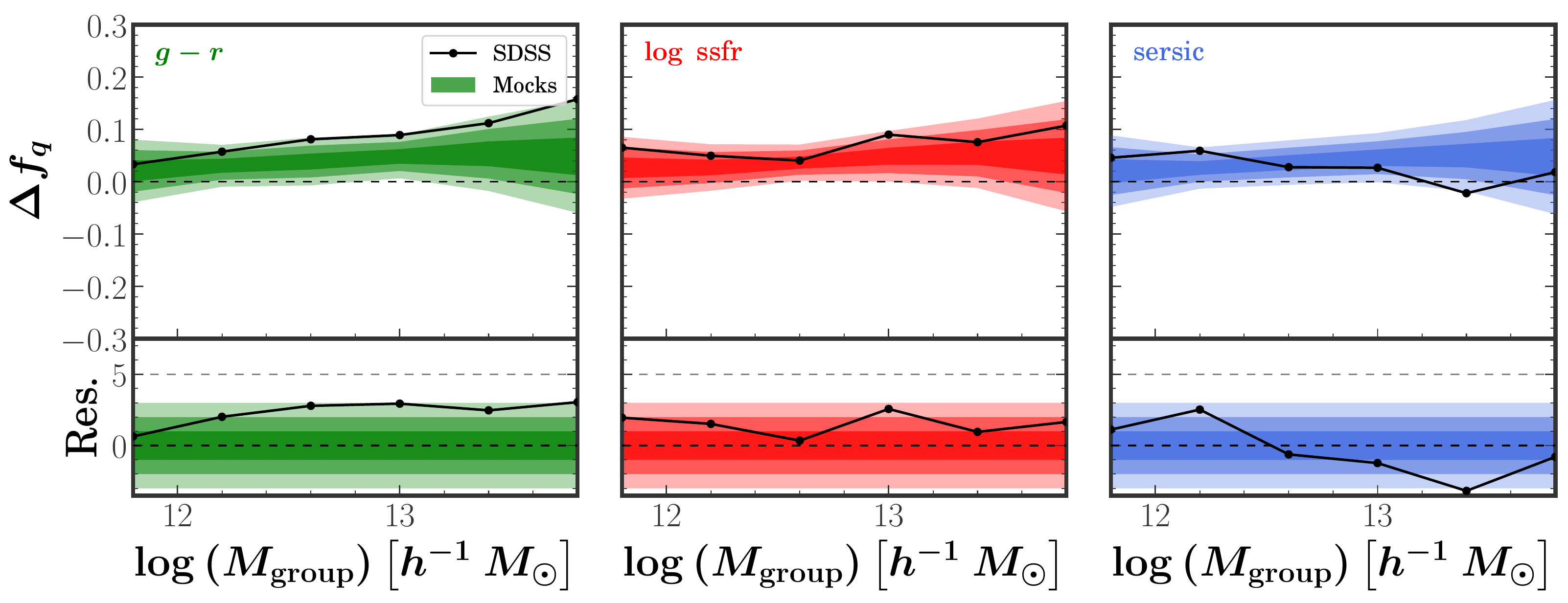}
    \end{centering}
    \caption{
    Similar to Figure~\ref{Fig:one_halo_fractions_data}, except that
    the \fracdiffm{q} of \MD{19}
    are compared to the
    distributions of measurements from mock catalogues rather
    than randomly shuffled data. \textit{Top panels}: The
    solid black lines correspond to the 
    \fracdiffm{q}
    in \MD{19}. The shaded contours show the \sig{1}, \sig{2},
    and \sig{3} ranges of \fracdiffm{q} calculated from
    100 mock catalogues with no built-in conformity.
    \textit{Bottom panel}: Normalised residuals of
    \fracdiffm{q} with respect to the mock catalogues.
    The solid black lines show the difference between
    \fracdiffm{q} for \MD{19} and the mean of the mocks,
    divided by the standard deviation of the mocks. The shaded
    contours show the \sig{1}, \sig{2}, and \sig{3} ranges of
    the mocks in this normalised space.}
    \label{Fig:one_halo_fractions_mocks}
\end{figure*}

Figure~\ref{Fig:one_halo_fractions_mocks} is analogous to 
Figure~\ref{Fig:one_halo_fractions_data}, except that the shaded contours now
show the distribution of mocks rather than shuffles. The black lines in the
top panels show the \fracdiffm{q} for each galaxy property as measured in
the \MD{19} sample and are thus identical to the black lines in the three
top panels of Figure~\ref{Fig:one_halo_fractions_data}. The shaded contours
show the \sig{1}, \sig{2}, and \sig{3} ranges of \fracdiffm{q,mock} values
of the mock catalogues. The bottom panels show the residuals with respect
to the mocks, as defined in equation~(\ref{eq:fgalresmock_eq}).
The prominent conformity signals that we found previously disappear
when compared against the mock catalogues. This is because the whole shaded
bands are no longer centred at $\fracdiff{q}=0$, but have shifted up
significantly. In other words, the mock catalogues with no built-in
conformity have an average quenched fraction difference of 0.02 to 0.05
for the three galaxy properties, depending on group mass. These
{\textit{spurious}} conformity signals must be due to grouping errors
-- either in misidentification of centrals and satellites, or in estimation
of \mgroup. We have examined the contributions to the induced signal from
both of these factors by constructing versions of our mock catalogues that do
not contain these errors. We find that most of the induced signal comes
from errors in assigning \mgroup, consistent with \citet{Campbell2015d}.

Figure~\ref{Fig:one_halo_fractions_mocks} shows that the statistical
significance of the 1-halo conformity signal in the quenched fractions of the
\MD{19} sample drops from $5-6\sigma$ when using the shuffles to $2.5-3\sigma$
when using the mocks.
Consequently, we can no longer claim a significant detection of 1-halo galactic
conformity. This result illustrates the importance of using mock catalogues to
compute the null model (i.e., no conformity case) in any conformity analysis.
Moreover, it is necessary to use a large suite of mock catalogues to properly
specify the distribution of the null model. A few of our 100 mock catalogues
do not display spurious conformity signals and so if we had only used one mock
that happened to lack any conformity signals, we would have come to the wrong
conclusion about the significance of our conformity detection. Our result
calls into question previous claims of 1-halo conformity detections,
especially from papers that used similar group-based methods as ours,
including the original detection by \citetalias{Weinmann2006}.

\subsubsection{\mcft for 1-halo Conformity}
\label{subsub:one_halo_mcf}

We now move to the second statistic that we are using to probe galactic
conformity, the ``marked correlation function'', \mcf. Since the \mcf can be
more sensitive than binary statistics, and can potentially uncover the
scale dependence of any correlations (see discussion in
\refsec{sub:marked_corr}), the \mcf is well-suited to exploring the
correlations between central and satellite galaxies.

We evaluate \mcf for the three galaxy properties, i.e. \gr colour, \ssfr,
and \sersic index, in different bins of \mgroup. Each galaxy pair is
comprised of a central galaxy and a satellite galaxy of the same galaxy group,
and the projected distance, $r_{p}$, is the distance between the two member
galaxies. We then take the product of the `marks' of the two galaxies and
average this over all pairs in bins of $r_{p}$. The mark for each galaxy
is just the value of its property (e.g., colour) normalised by the mean value
over the whole population of similar galaxies. We do this in two ways. First,
we normalise using the mean of all central or satellite galaxies in the same
bin of \mgroup. For example, the colour of each central (satellite) galaxy
is divided by the mean colour of all central (satellite) galaxies that live
in similar mass groups. \mcf then measures the correlation coefficient between
the normalised colours of central and satellite galaxies. Since this
measurement is done in $r_p$ bins, it is sensitive to radial gradients in the
properties of satellite galaxies within groups, typically referred to as
\textit{segregation}. For example, if groups contain colour segregation in the
sense that satellite galaxies in the central regions of groups tend to be
redder than satellite galaxies in the outskirts of groups, then \mcf will be
larger than unity in bins of small $r_p$ and less than unity in bins of
large $r_p$. Such a radial segregation effect will masquerade as a 1-halo
conformity signal. To account for this, we do a second normalization where the
properties of satellite galaxies are normalised by the mean values of all
satellites that live in the same bin of both \mgroup and $r_p$.
Measured in this way, \mcf is not sensitive to radial segregation and so
values different from unity are direct indications of conformity.

To assess the statistical significance of a conformity signal while at the
same time avoiding any biases due to grouping errors, we now only compare the
results of the \mcf of \MD{19} to those of the mock catalogues and not to
those from the shuffling technique. By making this type of comparison,
we avoid systematic errors that might masquerade as conformity signals. For 
example, it may be the case that galaxies that live in the outskirts of
large groups are more likely to have been mis-assigned to their group than
galaxies in the central regions of groups. These ``satellites'' may actually
be centrals in much smaller neighbouring haloes that were incorrectly merged
into the large groups. Since these low-mass centrals are likely to be bluer
in colour than actual satellites of the large group, this error will
masquerade as a radial colour gradient within groups.
Such an effect may represent itself as an anti-correlation
at large 1-halo scales. This type of systematic error will be present in
the mocks as well and so we can account for the role of grouping errors
by comparing our measurements to mock catalogues that contain no built-in
conformity or segregation, but are analyzed in the same way as the SDSS data

Like we did for the quenched fraction differences in 
\refsec{subsub:one_halo_fracs}, we analyse the 100 mock catalogues in the same
way as we analyse the \MD{19} data. Specifically, we compute \mcf of each
galaxy property, i.e. \ssfr, \gr colour, and \sersic index, on the mocks after
first normalizing each galaxy property the two different ways
(in bins of \mgroup and in bins for both \mgroup and \rp). We use the standard
deviation of \mcf values to estimate the \sig{1}, \sig{2}, and \sig{3} ranges
of the distributions for each galaxy property. We then determine the
statistical significance of the result by calculating the residuals of the
SDSS measurements with respect to mocks as
\begin{align}
\label{eq:mcf_residuals_mock}
\textrm{Res} = \frac{\mcfeq - \overline{\mcfeq_\mathrm{mock}}}{\sigma_\mathrm{mock}}.
\end{align}
This is similar to the residuals in equation~(\ref{eq:fgalresmock_eq}).

\begin{figure*}
    \centering
    \begin{centering}
    \centering
    \includegraphics[ height = 0.80 \textheight]{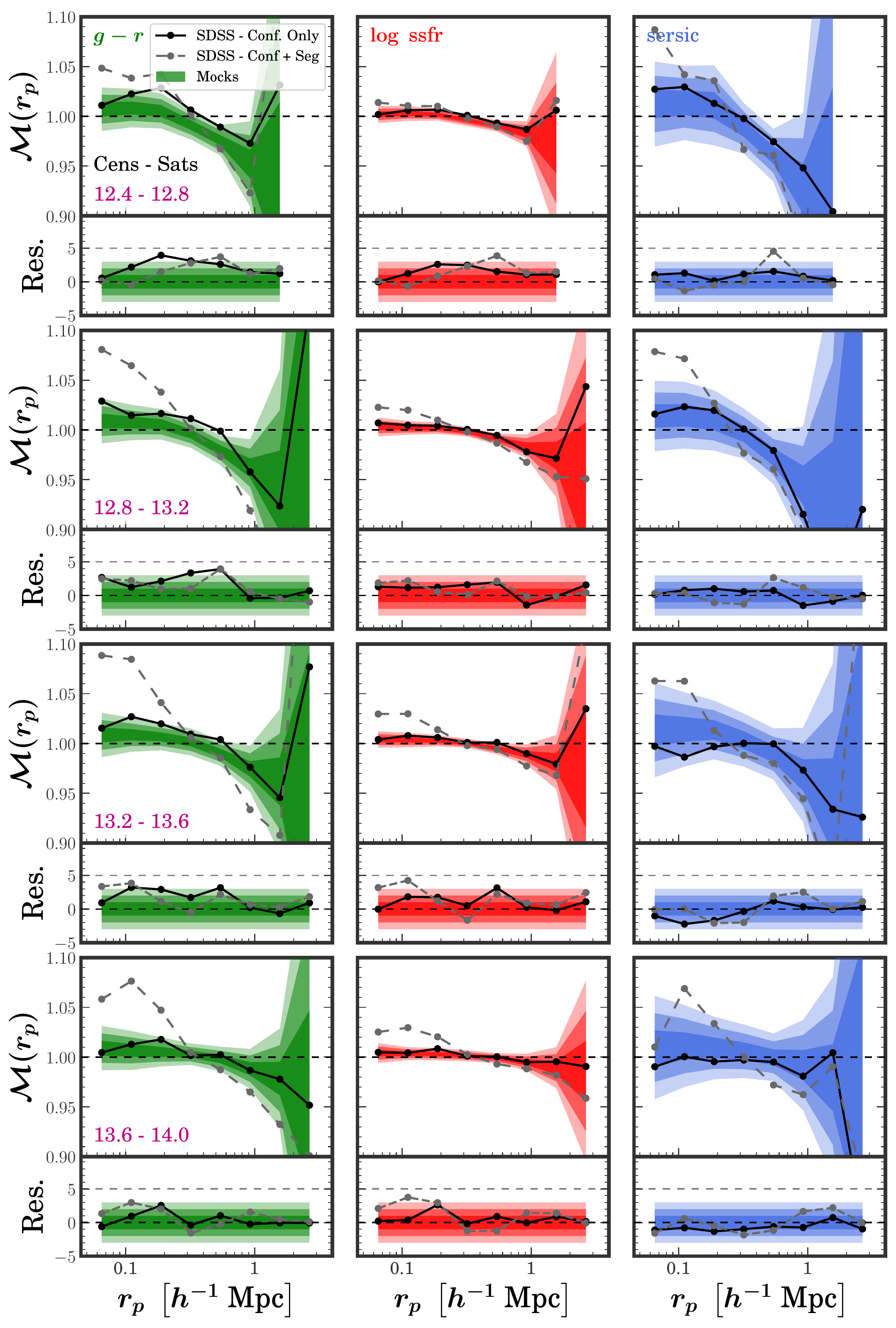}
    \end{centering}
    \caption{
        Marked correlation function, \mcf, of \gr colour (left),
        \ssfr (centre), and \sersic index (right), as a function
        of projected distance \rp for central-satellite galaxy
        pairs within the same galaxy groups in \MD{19} and mock catalogues.
        Each row corresponds to a bin of group mass, \mgroup, as
        listed in the left panels. \textit{Top panels}: The solid
        black lines correspond to the case where the marks have
        been normalised to remove the effects of radial segregation,
        while the dashed grey lines include segregation. The shaded
        contours show the \sig{1}, \sig{2}, and \sig{3} ranges of
        \mcf calculated from 100 mock catalogues with no built-in
        conformity or radial segregation. These mock results
        can only be compared to the solid black lines. We do not
        show the mock results that correspond to the dashed grey lines.
        \textit{Bottom panels}: Normalised residuals of \mcf with
        respect to the mock catalogues. The lines show the difference 
        between \mcf of the SDSS and the mean of the mocks, divided 
        by the standard deviation of \mcf for the mocks. The solid 
        black and dashed grey lines correspond to the cases where 
        effects of radial segregation are removed and included, 
        respectively. The shaded contours show the \sig{1}, \sig{2}, 
        and \sig{3} ranges of the mocks in this normalised space.}
    \label{Fig:one_halo_mcf_censat_mocks}
\end{figure*}

Figure~\ref{Fig:one_halo_mcf_censat_mocks} shows \mcf of \gr colour (left),
\ssfr (centre), and \sersic index (right), as a function of projected distance,
\rp, with each row corresponding to a bin of \mgroup. In this figure, we
only show bins with $\mgroup > 10^{12.4}\msunh$ since these exhibited the
largest signals in the quenched fraction difference statistic for colour and
\ssfr, as shown in Figure~\ref{Fig:one_halo_fractions_data}. In the top part
of each panel, the black, solid line corresponds to the \mcf of SDSS galaxies,
when properties are normalised within bins of \rp in order to remove the
effects of radial segregation. For comparison, the grey dashed line
corresponds to the case when the segregation effect is included, i.e.,
the contributions for the \mcf results are coming from both galactic
conformity and the segregation effect. The shaded regions correspond to the
\sig{1}, \sig{2}, and \sig{3} ranges of the distributions of \mcf values
for mock catalogues. However, these results are analysed by normalizing
properties within bins of \rp, so only the black, solid lines can be
compared to the shaded regions. We do not show the results that correspond
to the grey, dashed lines. The bottom part of  each panel shows the
residuals of each \mcf with respect to the mock catalogues, as defined
in equation~(\ref{eq:mcf_residuals_mock}). In this case, the black solid
lines and grey dashed lines are each computed using their corresponding
set of mock results.

In Figure~\ref{Fig:one_halo_mcf_censat_mocks} the shaded regions for
\gr colour, \ssfr, and \sersic index are not centred at \mcf=1,
indicating the effect of group errors. The strength of both radial segregation
and conformity signals in the SDSS are weak when compared to mock catalogues 
containing neither effect. First we examine the case where we normalise galaxy 
properties by their mean values in bins of \mgroup, making \mcf sensitive to
both conformity and segregation (dashed grey lines). We do detect significant
radial segregation (dashed grey lines) for colour and \ssfr at scales
smaller than $0.2\mpch$ in the case of massive groups, in the sense that
satellite galaxies close to the centres of their groups tend to be more
quenched (and thus more similar to their central galaxies) than satellite
galaxies farther out. We do not find such correlations for the
\sersic index at those scales. Next we examine the case where radial 
segregation is removed (solid black lines). The 1-halo conformity signal 
hovers near the \sig{3} level for a wide range of small scales
for colour and \ssfr. However, the signal is not strong enough for us to claim
a conformity detection. In summary, neither the quenched
fractions nor the marked correlation function reveal any statistically
significant 1-halo conformity signal after controlling for group errors
for the cases of \gr colour, \ssfr, and \sersic index.

\subsection{2-halo Conformity}
\label{sub:two_halo_conf}

We next study 2-halo conformity, which is the correlation of properties for
galaxies that live in separate haloes. As we discussed in \refsec{sec:intro},
a detection of 2-halo conformity in \ssfr was claimed by
\citetalias{Kauffmann2013a} for low-mass central galaxies out to scales of 4
Mpc. This claim has been challenged by \citetalias{Tinker2017a} and
\citetalias{Sin2017} who reproduced the result of \citetalias{Kauffmann2013a}
and showed that the conformity signal is mainly driven by contamination in
the isolation criterion to select the sample of central galaxies.
After removing a small fraction of satellite galaxies that were
misclassified as centrals from the primary sample, only a weak conformity
signal remains out to projected distances of 2 Mpc.

\subsubsection{Quenched fractions for 2-halo Conformity}
\label{subsub:two_halo_conf_quench_frac}

We first analyse \twoh conformity using the quenched fraction difference
statistic, which is similar to what was used by some of these previous works.
We use a sample composed of only central galaxies as classified by our
group-finder, which are the most luminous galaxies in the $r$-band within
their respective groups. Then, using the same group mass bins as before,
we compute the quenched fraction difference statistic, as described in
\refsec{sub:quenched_frac_method}, in bins of projected separation \rp and
only counting galaxy pairs within a line-of-sight separation of
$\pi_\mathrm{max}=20\mpch$. For example, to calculate \fracdiffm{red} for the
smallest group mass bin we consider, we first list all the central galaxies
in groups with log\mgroup: 11.6--12.0, then find all pairs of these galaxies
that have line-of-sight separations less than $\pi_\mathrm{max}$, and place
them in logarithmic bins of $r_p$. Each radial bin now contains a set of
central-central galaxy pairs where one of the galaxies is designated as
``primary" and the other as ``secondary" (each pair is counted twice so that
both galaxies have a turn at being primary). We then make one list of pairs
where the primary is red and another where it is blue. For each list we then
calculate the fraction of pairs where the secondary is red (i.e., the
``quenched fraction") and we take the difference between these two fractions.
We repeat this procedure for all group mass bins and for \ssfr and \sersic
index. As before, we assess the statistical significance of conformity
signals by comparing with our set of 100 mock catalogues that contain no 
intrinsic conformity, but do contain the same types of systematic errors that
affect the SDSS analysis.

\begin{figure*}
    \centering
    \includegraphics[ height = 0.785 \textheight ]{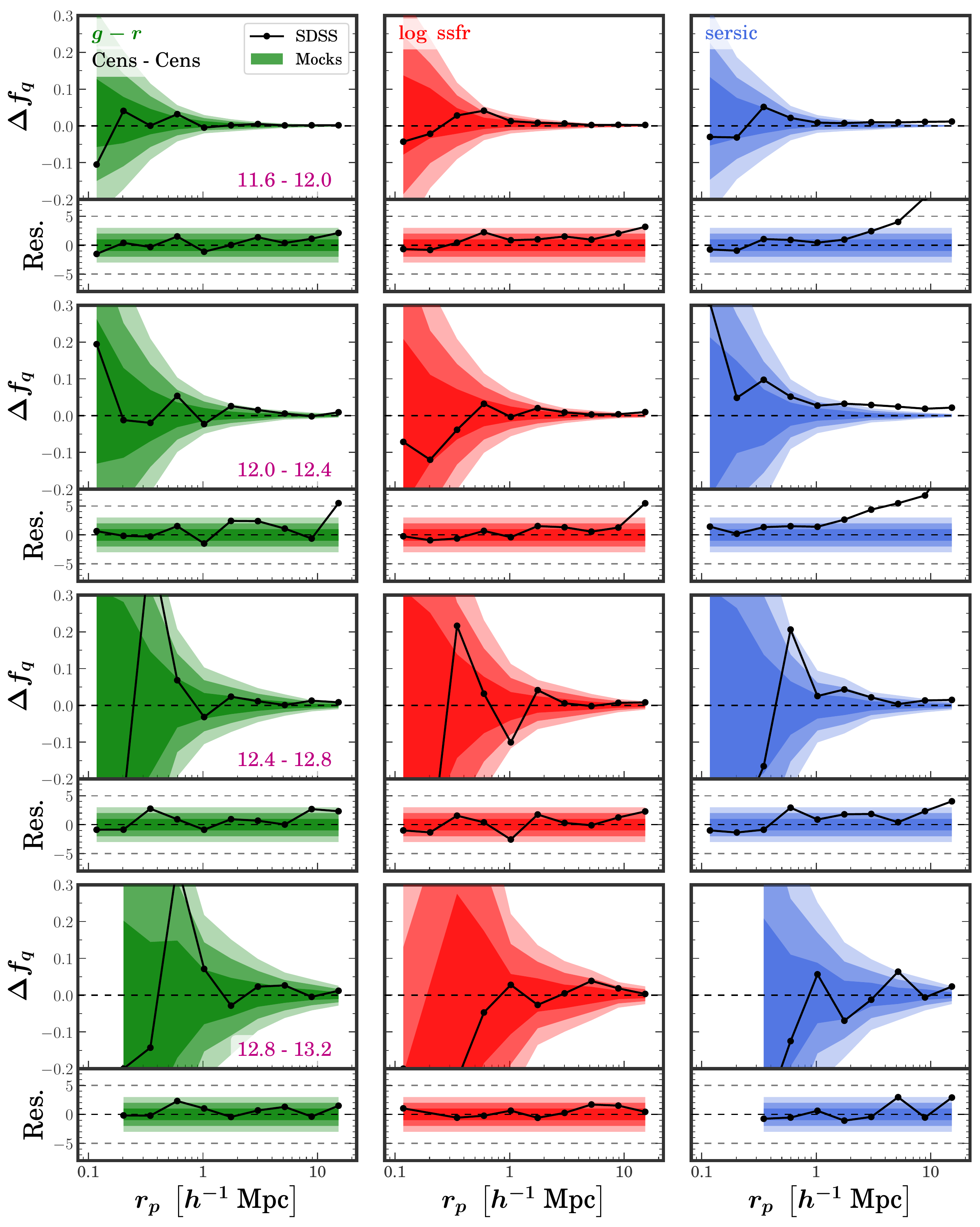}
    \caption{
    Difference of fractions, \fracdiffm{}, of red (left),
    passive (centre), and early-type (right) secondary
    central galaxies as a function of their projected distance,
    \rp, from primary central galaxies in groups of similar mass,
    where the difference is measured between primary galaxies
    that are red and blue, passive and active, early-type and
    late-type, respectively. Each row corresponds to a bin
    of group mass, \mgroup, as listed in the left panels.
    \textit{Top panels}: The solid black lines correspond to
    the \fracdiffm{} of each galaxy property in \MD{19}. The 
    shaded contours show the \sig{1}, \sig{2}, and \sig{3} ranges of
    \fracdiffm{} calculated from 100 mock catalogues
    with no built-in conformity. \textit{Bottom panels}:
    Normalised residuals of \fracdiffm{} with respect
    to the mock catalogues. The solid black lines show the
    difference between \fracdiffm{} for \MD{19} and
    the mean of the mocks, divided by the standard deviation
    of \fracdiffm{} for the mocks. The shaded contours show the \sig{1},
    \sig{2}, and \sig{3} ranges of the mocks in this normalised
    space.
    }
    \label{Fig:two_halo_fracs_cen_mocks}
\end{figure*}


Figure~\ref{Fig:two_halo_fracs_cen_mocks} presents our main results of
probing 2-halo conformity using quenched fraction differences. The three
columns show results for \gr colour (left column), \ssfr (middle column),
and \sersic index (right column), as measured in the \MD{19} sample and mock
catalogues. Each row corresponds to a bin of \mgroup, as listed in the left
column of plots. We focus on the four lowest-mass bins since
\citetalias{Kauffmann2013a} found 2-halo conformity signals at these
masses. The black lines in the top portions of each panel show the
\fracdiffm{q} as a function of projected separation \rp, while the shaded
contours show the \sig{1}, \sig{2}, and \sig{3} ranges of \fracdiffm{q, mock}
for the 100 mock catalogues of each galaxy property. The bottom panels
show the residuals of each galaxy property with respect to the mock
catalogues, as defined in equation~(\ref{eq:fgalresmock_eq}).


Figure~\ref{Fig:two_halo_fracs_cen_mocks} does not reveal any 2-halo
conformity signals for most group masses and scales for \gr colour and
\ssfr. \sersic index exhibits a prominent 2-halo conformity signal for the
two lowest-mass bins, i.e., for group masses of log\mgroup 11.6--12.4 and
at scales of $r_{p}>3\mpch$. This large \sersic index signal is caused by
the fact that SDSS central galaxies in groups of log\mgroup 11.6--12.4
exhibit a small \fracdiffm{early}=1--2\% that is constant with scale, while
the scatter among the mock catalogues reduces with scale, resulting in a
strongly increasing significance of the conformity signal. This figure
also shows that the mock results are perfectly centred at \fracdiffm{q}=0,
which means that group errors do not seem to impact 2-halo conformity
measurements nearly as much as they did in the 1-halo case.

\subsubsection{\mcft for 2-halo Conformity}
\label{subsub:two_halo_conf_mcf}

We next study 2-halo conformity using the marked correlation function \mcf.
We perform a similar analysis as the 1-halo case presented in
\refsec{subsub:one_halo_mcf}, except that now we only consider pairs of
central galaxies from different groups of similar mass. As with the quenched
fraction difference case, we count all central-central pairs with a
line-of-sight separation less than $\pi_\mathrm{max}=20\hmpc$ and place
them in logarithmic bins of projected distance \rp.
We then compute \mcf for our three galaxy properties after normalizing them
by their mean values within \mgroup bins.

To assess the statistical significance of our results and investigate
the impact of grouping errors and mass assignment, we compare our
SDSS results with measurements on our 100 mock catalogues that contain
no built-in 2-halo conformity. Once again, we use the standard deviation
of mock \mcf values to estimate the \sig{1}, \sig{2}, and \sig{3} ranges of
the mock distribution. We then calculate the residuals of the SDSS measurements
with respect to mocks as in equation~(\ref{eq:mcf_residuals_mock}).

\begin{figure*}
    \centering
    \includegraphics[height = 0.80 \textheight]{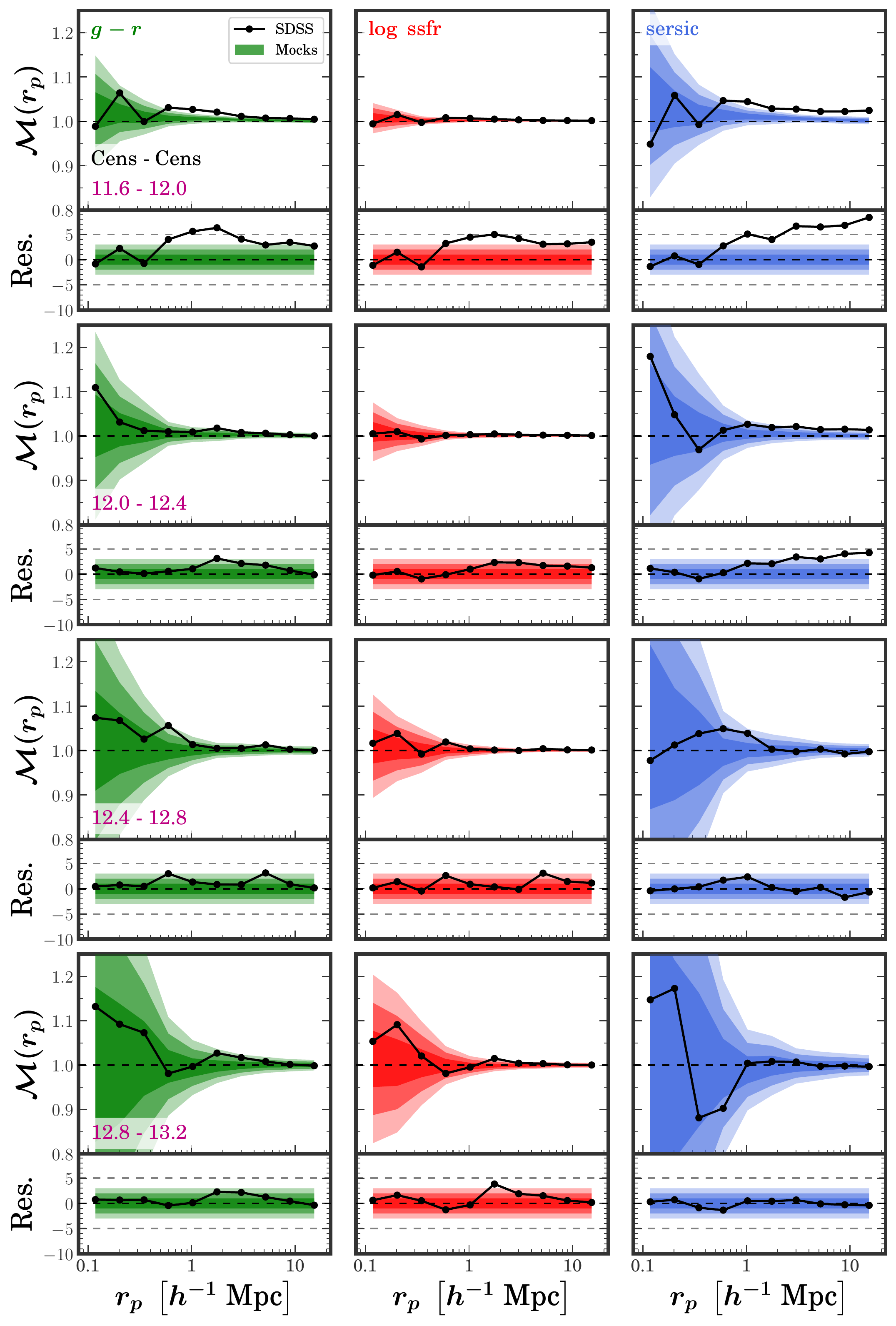}
    \caption{
        Mark correlation function, \mcf, of \gr colour (left),
        \ssfr (centre), and \sersic index (right), as a function
        of projected distance \rp for central-central
        galaxy pairs within separate galaxy groups in the \MD{19} sample and
        mock catalogues. Each row corresponds to a bin of group mass,
        \mgroup, as listed in the left panels.
        \textit{Top panels}: The solid black lines show results for SDSS,
        while the shaded contours show the \sig{1}, \sig{2}, and \sig{3}
        ranges of \mcf calculated from 100 mock catalogues with no
        built in 2-halo conformity. \textit{Bottom panels}:
        Normalised residuals of \mcf with respect to the mock
        catalogues. The solid black lines show the difference
        between \mcf of the SDSS and the mean of the mocks,
        divided by the standard deviation of \mcf for the
        mocks. The shaded contours show the \sig{1}, \sig{2}, and
        \sig{3} ranges of the mocks in this normalised space. 
    }
    \label{Fig:two_halo_mcf_cencen_mocks}
\end{figure*}

Figure~\ref{Fig:two_halo_mcf_cencen_mocks} shows the \mcf of
\gr colour (left), \ssfr (middle), and \sersic index (right), as a function
of projected distance, \rp, with each row corresponding to a bin of \mgroup.
They layout is similar to that in the previous figures. The figure reveals
weak, but highly significant 2-halo conformity signals for all three
properties in low mass haloes. In the lowest mass bin, log\mgroup: 11.6--12.0,
these signals reach a significance as high as \sig{7}.
In the case of \gr colour, the signal reaches as high as \mcf=1.02-1.03 and
then declines with scale, while the statistical significance peaks at scales
$r_{p}: 0.6-4\hmpc$ and hovers at the \sig{3} level out to $r_p\sim 10\hmpc$
before dropping at larger scales. There is no significant large-scale
conformity signal in more massive group bins. \ssfr behaves the same way,
except that the conformity signal is much weaker (yet equally signifiant), 
peaking at \mcf value of less than 1.007. In the case of \sersic index, 
the signal is also as high as \mcf=1.02-1.03, but, unlike with colour, it 
keeps this constant amplitude out to the largest scales we consider. As a 
result, the statistical significance of the conformity signal keeps rising 
with scale because the scatter in the mock distribution decreases with scale. 
In the next mass bin, log\mgroup: 12.0-12.4, the conformity signals almost 
disappear, but are still significant for \sersic index. There are no 2-halo 
conformity signals in the higher group mass bins.

These results are
similar to what we found using the quenched fraction difference statistic,
where \sersic index displayed the strongest 2-halo conformity signal but
only for the low mass groups. However, the marked correlation function is
a more sensitive statistic for detecting 2-halo conformity as demonstrated
by the much higher statistical significance of the weak observed signals in
the case of color and \ssfr. Where we found no strong evidence of 2-halo 
conformity using \fracdiffm{} in Figure~\ref{Fig:two_halo_fracs_cen_mocks}, 
we find strong such evidence using \mcf in Figure~\ref{Fig:two_halo_mcf_cencen_mocks}. 
The marked correlation function is clearly a more sensitive probe of 2-halo 
conformity than quenched fractions, and gives us a better handle on 2-halo 
conformity signals for colour and \ssfr.
In summary, we have found low amplitude, but highly significant \twoh
conformity signals for \gr colour and \ssfr out to $4\hmpc$ and an intriguing
signal in \sersic index out to the largest scales that we probe.

\section{Summary and Discussion}\label{sec:summary_discussion}

In this paper, we study galactic conformity, which is the phenomenon that
galaxy properties, such as colour or morphology, may exhibit correlations
across distance, beyond what would be expected if these properties only
depended on halo mass. At small scales, this ``1-halo conformity" is seen
as a correlation between the properties of satellite galaxies with those of
the central galaxy whose halo they inhabit. At large scales,
``2-halo conformity" is seen as a correlation between central galaxies in
haloes that are well separated from each other. In both cases, it is
important to control for halo mass in order to ensure that any detected
correlations are not simply due to the well-established correlations between
galaxy properties and halo mass, as well as the correlation between halo
mass and larger-scale environment. We are motivated to perform a comprehensive
study of conformity because recent works have
exposed systematic problems with previous claims of conformity detection at
$z=0$, calling into question whether conformity has actually been detected.
In the 1-halo regime, the original detection came from
\citetalias{Weinmann2006} using a group catalogue to designate central and
satellite galaxies and to control for halo mass. However,
\citet{Campbell2015} used a mock catalogue to show that errors in
group-finding and group mass assignment can lead to a spurious 1-halo
conformity signal when none is actually present. In the 2-halo regime,
\citetalias{Kauffmann2013a} detected conformity out to 4 Mpc using isolation
criteria to avoid including satellite galaxies. However,
\citetalias{Tinker2017a} and \citetalias{Sin2017} showed that this
result was most likely due to insufficiently stringent isolation criteria
and that the detected conformity signal arose from a small number of
satellite galaxies that were misidentified as centrals.

We investigate both 1-halo and 2-halo conformity using a galaxy group
catalogue from the SDSS DR7. Our analysis contains three main improvements
over previous works. First, we study three observed properties of galaxies:
\gr colour, \ssfr, and \sersic index. Second, we use a new statistic, the
marked correlation function, \mcf, in addition to the previously used
quenched fractions. \mcf is ideally suited for conformity studies
and is a more sensitive probe of weak conformity signals. Third, we use a
suite of 100 mock galaxy catalogues to quantify the statistical significance
of our results. These mock catalogues have the same clustering and same
distributions of ``observed" properties as the SDSS data
(luminosity, \gr colour, \ssfr, and \sersic index), and we analyse them in
exactly the same way (i.e., same group-finding algorithm, same way of
assigning group masses, etc). The mock catalogues do not have any built-in
conformity, but they are affected by the same systematic errors as the SDSS
data. By comparing our SDSS measurements to the distribution of mock
measurements, we can quantify the probability that whatever signal we
detect could have arisen from a model with no conformity.

The main results of our work are as follows.
\begin{itemize}[leftmargin=0.5\parindent, labelsep=0.5\parindent]
\item
When measuring the difference between quenched fractions of satellite galaxies
around quenched vs. non-quenched centrals, we detect a strong 1-halo
conformity signal at all group masses, which is strongest for \gr colour,
somewhat weaker for \ssfr, and only significant at low masses for \sersic
index. These results are in perfect agreement with the results of
\citetalias{Weinmann2006}. However, when we compare the \gr colour, \ssfr,
and \sersic index results to measurements made on our mock catalogues,
we find that they are also in perfect agreement. Since the mock catalogues 
contain no built-in conformity, this strongly suggests that the conformity
signal we detected is a result of systematic errors in the group mass
estimation and in central/satellite mis-assignment. This calls into
question the validity of the \citetalias{Weinmann2006} detection, as well as
other 1-halo conformity detections at $z=0$ that use group catalogues.

\item
The marked correlation function, \mcf, calculated with central-satellite
galaxy pairs is sensitive to the radial segregation of satellite galaxy
properties within groups. Using the 1-halo \mcf, we find significant
radial segregation for colour and \ssfr at scales smaller than $r_p<0.2\mpch$
in the case of groups more massive than log\mgroup>13. We do not find such
a signal for \sersic index. We thus claim a detection of radial segregation
in \gr color and \ssfr.

\item
After removing the effect of radial segregation from \mcf by properly
renormalising galaxy properties, the amplitude of \mcf reduces and
the conformity signal mostly vanishes. We thus do not detect 1-halo 
conformity using the \mcf statistic in any of the three galaxy
properties.

\item
Studying the quenched fraction difference statistic as a function of projected
scale for central-central galaxy pairs in groups of similar mass reveals
no 2-halo conformity signal for \gr colour or \ssfr. However, we find a
highly significant 2-halo conformity signal for \sersic index in low
mass groups of log\mgroup < 12.4. This signal is constant with scale and
thus increases in statistical significance with scale.
The mock measurements of the three galaxy properties indicate that group
errors do not strongly affect our 2-halo quenched fractions, and that
the detection of 2-halo conformity in \sersic index is likely robust.

\item
The \mcf of central-central galaxy pairs proves to be a more sensitive
probe of conformity than quenched fractions. We find a low amplitude, yet 
highly significant signal in all three galaxy properties for group masses 
below log\mgroup=12. For \gr colour and \ssfr, the signal is strongest at 
scales of $r_p: 0.6-4\hmpc$ and hovers at the \sig{3} level out to
$r_p\sim 10\hmpc$ before dropping at larger scales. For \sersic index,
the 2-halo conformity signal increases in significance with scale.
There is no significant large-scale conformity signal in more massive groups.
Our detection is unlikely caused by group errors and thus represents robust
2-halo conformity detections in colour, \ssfr, and \sersic index for
central-central galaxy pairs at low masses.

\end{itemize}

These results demonstrate the importance of using mock galaxy catalogues
in any study of galactic conformity. Comparing our SDSS measurements with
the distribution of mock measurements allows us to test the null model
(i.e., no conformity) in a way that includes systematic errors in
group-finding or mass estimation. Without the mock catalogues, we would
have claimed a strong detection of 1-halo conformity. Instead, we
are driven to the conclusion that the 1-halo signal is not real.
This result calls into question whether any study has actually detected
1-halo conformity in the SDSS data. The one caveat to these conclusions
is that they only hold to the extent that our mock catalogues faithfully
represent the real universe. If, for example, the correlation between
\ssfr and halo mass in the mocks is not as strong as it should be,
then the impact of group mass errors on the conformity signal will
not be accurate.

In the case of 2-halo conformity, we do not find any statistically
significant signals when looking at quenched fractions using colour or
\ssfr. We thus agree with the claim in \citetalias{Tinker2017a}, that
the \citetalias{Kauffmann2013a} result must have suffered from errors
in the isolation criteria used. On the other hand, we show that
the marked correlation function is more sensitive to the underlying weak
signal and displays a clear conformity trend, even when compared against
the mock catalogues. This measurement may thus represent the first robust
detection of 2-halo conformity to-date. Our finding that 2-halo
conformity is strongest when considering galaxy \sersic index
is curious and merits further study. Overall, to understand the
physical origin of these conformity signals, it will be necessary to
model them in detail, which we leave for future work.

\section{Acknowledgements}\label{sec:acknowledments}

The mock catalogues used in this paper were produced by the LasDamas project
(\url{http://lss.phy.vanderbilt.edu/lasdamas/}); we thank NSF XSEDE for
providing the computational resources for LasDamas. Some of the
computational facilities used in this project were provided by the
Vanderbilt Advanced Computing Center for Research and Education (ACCRE).
This project has been supported by the National Science Foundation (NSF)
through a Career Award (AST-1151650).
Parts of this research were conducted by the Australian Research Council
Centre of Excellence for All Sky Astrophysics in 3 Dimensions (ASTRO 3D),
through project number CE170100013.
This research has made use of
NASA's Astrophysics Data System. This work made use of the IPython package
\citep{PER-GRA:2007}, Scikit-learn \citep{mckinney}, SciPy
\citep{jones_scipy_2001}, matplotlib, a Python library for publication
quality graphics \citep{Hunter:2007}, Astropy, a community-developed
core Python package for Astronomy \citep{2013AA...558A..33A}, and
NumPy \citep{van2011numpy}. Funding for the SDSS and SDSS-II has been
provided by the Alfred P. Sloan Foundation, the Participating Institutions,
the National Science Foundation, the U.S. Department of Energy,
the National Aeronautics and Space Administration, the
Japanese Monbukagakusho, the Max Planck Society, and the Higher Education
Funding Council for England. The SDSS Web Site is http://www.sdss.org/.
The SDSS is managed by the Astrophysical Research Consortium for the
Participating Institutions. The Participating Institutions are the
American Museum of Natural History, Astrophysical Institute Potsdam,
University of Basel, University of Cambridge, Case Western Reserve
University, University of Chicago, Drexel University, Fermilab, the
Institute for Advanced Study, the Japan Participation Group,
Johns Hopkins University, the Joint Institute for Nuclear Astrophysics,
the Kavli Institute for Particle Astrophysics and Cosmology,
the Korean Scientist Group, the Chinese Academy of Sciences (LAMOST),
Los Alamos National Laboratory, the Max-Planck-Institute for Astronomy (MPIA),
the Max-Planck-Institute for Astrophysics (MPA),
New Mexico State University, Ohio State University,
University of Pittsburgh, University of Portsmouth, Princeton University,
the United States Naval Observatory, and the University of Washington.
These acknowledgements were compiled using the Astronomy Acknowledgement
Generator.

\bibliographystyle{mnras}
\bibliography{Mendeley}

\bsp    
\label{lastpage}
\end{document}